\documentclass[journal]{IEEEtran}

\ifCLASSINFOpdf
\else
\fi

\usepackage{cite}

\usepackage[cmex10]{amsmath}

\usepackage[caption=false,font=footnotesize]{subfig}

\usepackage{stfloats}

\usepackage{graphicx}
\usepackage{color}
\usepackage{placeins}
\usepackage{float}
\usepackage[colorlinks=true,linkcolor=black,citecolor=black,urlcolor=black]{hyperref}
\usepackage{tabularx,colortbl,multirow}
\usepackage{enumerate}
\usepackage{enumitem}

\usepackage{bm}

\renewcommand\eqref[1]{(\ref{#1})}

\usepackage{tikz}
\usetikzlibrary{shapes,decorations,decorations.pathreplacing}
\usetikzlibrary{arrows,shapes,positioning,calc,fadings,patterns}
\usetikzlibrary{decorations,decorations.pathreplacing,decorations.pathmorphing}
\usepackage{tikz-3dplot}

\usepackage{booktabs}
\usepackage{threeparttable}
\usepackage{amssymb}

\usepackage{algorithm}
\usepackage{algcompatible}
\usepackage{amsmath}

\begin{document}
	
	\title{Physics-Constrained Inc-GAN for Tunnel Propagation Modeling from Sparse Line Measurements}
	
	\author{Yang Zhou, Haochang Wu, Yunxi Mu, Hao Qin, Xinyue Zhang, and~Xingqi~Zhang,~\IEEEmembership{Senior Member,~IEEE}
		
	\thanks{Manuscript received XX XX, XXXX; revised XX XX, XXXX; accepted XX XX, XXXX. \textit{(Corresponding author: Hao Qin)}}
	\thanks{Yang Zhou and Hao Qin are with the School of Electronics and Information Engineering, Sichuan University, Chengdu, China.}
	\thanks{Haochang Wu, Hao Qin, and Xinyue Zhang are with the School of Electrical and Electronic Engineering, University College Dublin, Dublin, Ireland.}
	\thanks{Yunxi Mu is with the College of Engineering, Peking University, Beijing, China.}
	\thanks{Xingqi Zhang is with the Department of Electrical and Computer Engineering, University of Alberta, Canada T6G 2H5.}
	\thanks{Digital Object Identifier: XXXX}
	
}
	
	\vspace{-1cm}
	
	\markboth{}%
	{Qin \textit{et al.}: PHYSICS-CONSTRAINED INCEPTION GAN FOR TUNNEL ELECTROMAGNETIC FIELD RECONSTRUCTION}
	
	\maketitle
	
	\begin{abstract}
		High-speed railway tunnel communication systems require reliable radio wave propagation prediction to ensure operational safety. However, conventional simulation methods face challenges of high computational complexity and inability to effectively process sparse measurement data collected during actual railway operations. This letter proposes an inception-enhanced generative adversarial network (Inc-GAN) that can reconstruct complete electric field distributions across tunnel cross-sections using sparse value lines measured during actual train operations as input. This directly addresses practical railway measurement constraints. Through an inception-based generator architecture and progressive training strategy, the method achieves robust reconstruction from single measurement signal lines to complete field distributions. Numerical simulation validation demonstrates that Inc-GAN can accurately predict electric fields based on measured data collected during actual train operations, with significantly improved computational efficiency compared to traditional methods, providing a novel solution for railway communication system optimization based on real operational data.
	\end{abstract}
	
	
	\IEEEpeerreviewmaketitle
	
	\section{Introduction}
	
	Modern high-speed railway systems represent critical infrastructure requiring robust and reliable wireless communication networks for train control, passenger services, and safety monitoring systems\cite{Qin25UAV, CHEN202578, qin2023physics}. In tunnel environments, electromagnetic (EM) field prediction becomes particularly challenging due to complex guided wave propagation modes, multiple boundary reflections, frequency dispersion effects, and strong modal coupling phenomena caused by confined geometries. These challenges are exacerbated by severe limitations faced in sensor deployment for high-speed trains, which typically can only carry single EM sensors for forward signal strength measurements\cite{Zhou17Alternative}\cite{He18Wave}. This constraint stems from space limitations, installation costs, safety regulations, and the dynamic nature of high-speed railway operations, making comprehensive field mapping equipment impractical.
	
	\begin{figure}[!ht]
		\centering
		\includegraphics[width=0.4\textwidth]{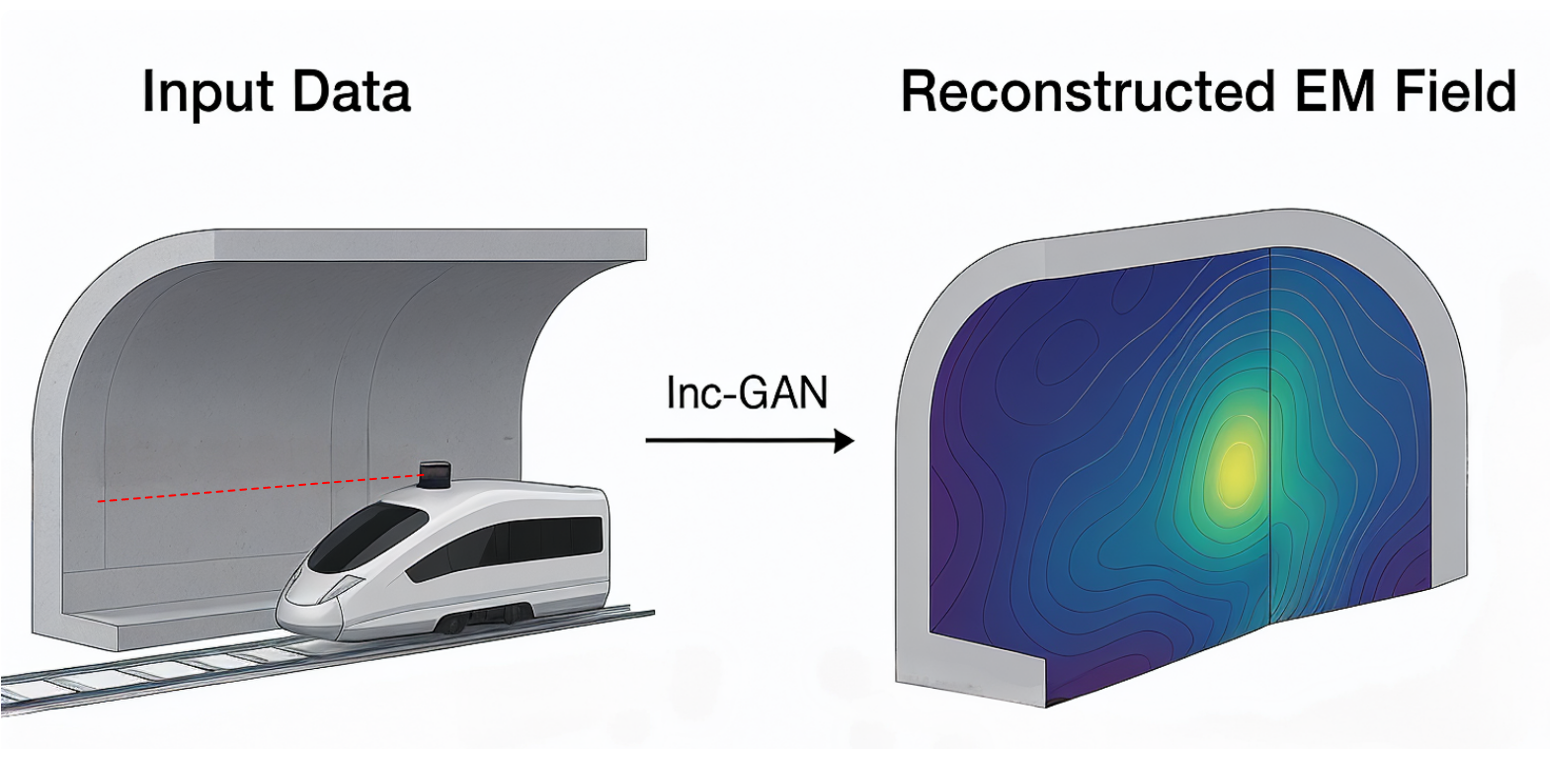}
		\caption{High-speed rail uses one line in a tunnel to predict electric field distribution across a plane.}
		\label{fig:introduction}
	\end{figure}
Accurate electric field prediction in railway tunnels is critical for real-time communication optimization and safety-critical connectivity. Railway operators require efficient channel quality assessment using sparse operational data to predict interference zones and optimize network performance. Traditional methods including finite-difference time-domain (FDTD)\cite{Yee66, Taflove05, Qian2025} and parabolic wave equation (PWE)\cite{Qin23_AWPL_SSSPE, Qin23comparative, Levy00, Qin25ToA}, while accurate, demand substantial computational resources and complete environmental characterization, making them unsuitable for processing sparse train measurement data or meeting real-time operational requirements.

	Recent years have witnessed promising developments in deep learning for wave propagation prediction~\cite{qin2023high, Seretis22, Seretis2020}. Terrain-aware propagation networks (TPN) have demonstrated effective prediction capabilities for irregular terrain \cite{Huang2025}. Artificial neural networks (ANN) have been successfully applied to tunnel electric field prediction. Conditional generative adversarial networks (CGAN) have addressed electric field distribution prediction problems based on transmitter characteristics \cite{Mallik2022}. Generative adversarial networks (GAN) variants have shown effectiveness in near-field channel estimation for massive MIMO systems\cite{Ye24GAN, Laci25Deep, Bordbar2024}.
	
	However, existing approaches have fundamental limitations: they cannot effectively utilize actual test data collected during train operations. Convolutional neural network (CNN)-based models require complete terrain profiles and detailed antenna parameters, while CGANs need precise transmitter location and antenna pattern information—none of which represent the type of test data available during actual train operations. More importantly, existing methods lack the capability to convert single signal line data measured by trains into complete electric field distributions, severely limiting their application value in practical operational environments.
	
The key innovation of Inc-GAN (Inception-enhanced generative adversarial network) lies in directly processing maximum energy value lines measured during actual train operations to reconstruct complete tunnel cross-section electric field distributions. Unlike existing methods that rely on theoretical assumptions or simulation data, our approach achieves sparse-to-dense electric field reconstruction using real operational data through an Inception-enhanced U-Net generator with physics-informed loss functions. The main contributions include: (1) a GAN capable of processing sparse operational measurements to reconstruct complete electric field distributions, (2) physics-informed loss formulations designed for handling actual measurement data while maintaining physical consistency, and (3) a multi-scale architecture optimized for tunnel electric propagation characteristics. This research bridges the gap between theoretical simulation and practical railway operations, providing the first solution for real-time electric field prediction using actual train measurement data.

	The rest of this letter is organized as follows.
	In Section \ref{PROPOSED INC-GAN RESTORATION MODEL}, we present our proposed Inc-GAN architecture including the multi-scale Inception blocks, generator and discriminator designs, and physics-informed loss functions.
	The numerical results and comprehensive performance evaluation of our approach are given in Section \ref{NUMERICAL RESULTS}, followed by practical applications in Section \ref{APPLICATION:}.
	Finally, we conclude the letter in Section \ref{CONCLUSION}.

	\section{PROPOSED INC-GAN RESTORATION MODEL}
	\label{PROPOSED INC-GAN RESTORATION MODEL}
	
	\subsection{Parabolic Wave Equation (PWE) Method}
	
	The PWE method is widely applied for radio wave propagation modeling in tunnel environments due to its favorable balance between accuracy and computational efficiency. This method discretizes the 3D computational domain $X \times Y \times Z$ into regular grids.
	
	The standard parabolic wave equation \cite{Levy00} can be expressed as:
	\begin{equation}
		\frac{\partial u}{\partial z} = \frac{1}{2jk_0}\left(\frac{\partial^2}{\partial x^2} + \frac{\partial^2}{\partial y^2}\right)u
		\label{eq:PE_standard}
	\end{equation}
	
	where $u$ represents the simplified plane wave solution and $k_0$ is the free-space wave number. However, as railway communication systems migrate toward higher frequency bands, the PWE method requires smaller discretization steps to ensure numerical accuracy, leading to dramatically increased computational costs. This challenge motivates our search for new methods capable of efficient electric field prediction using actual train measurement data.
	
	\subsection{Proposed Inc-GAN Architecture}
	Standard CGANs exhibit fundamental limitations for tunnel electromagnetic field reconstruction tasks \cite{Mirza14CGAN}. The fixed 3×3 convolutional kernels in traditional CGAN architectures fail to capture multi-scale spatial dependencies inherent in wave propagation. To address these multi-scale modeling limitations, we propose Inc-GAN, which employs multi-scale Inception blocks to capture both global and local electric field variations simultaneously.
	
	The framework of the proposed Inc-GAN electric field reconstruction model is as follows. During the training phase, a PWE solver generates complete tunnel cross-section electric field distributions as training datasets. Inc-GAN uses sparse signal lines actually measured by trains (signal source location rows) as input and performs end-to-end training with complete electric field distributions generated by the PWE solver as training targets. In practical application phases, the model requires only maximum energy value lines measured by train sensors as input, eliminating the need for complex PWE calculations, thus minimizing computational overhead and enabling real-time applications. Subsequently, Inc-GAN utilizes this sparse measured data to rapidly reconstruct complete tunnel cross-section electric field distributions, providing critical field strength information for real-time optimization and fault prediction of railway communication systems.
	
	\begin{figure}[!t]  
		\centering
		\includegraphics[width=0.55\textwidth]{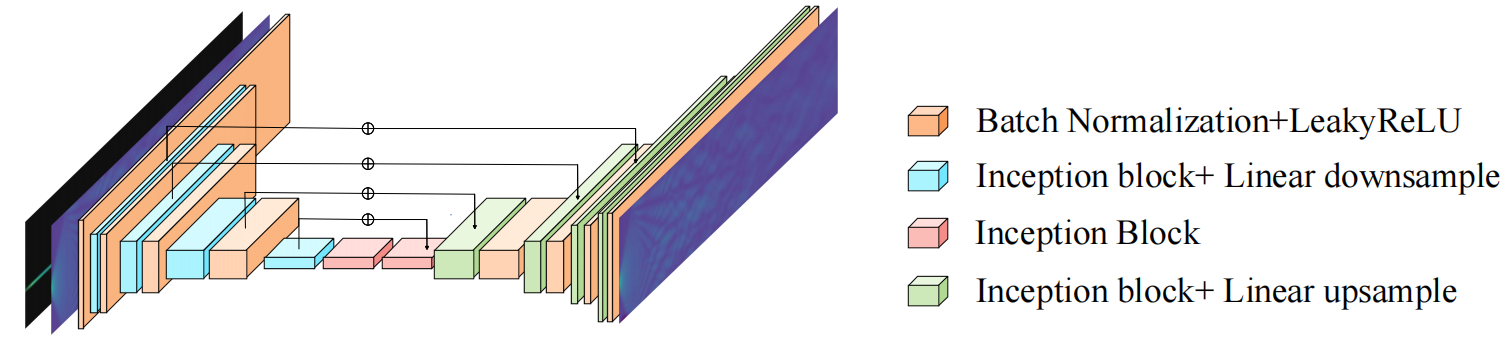}
		\caption{Inc-GAN generator architecture.}
		\label{fig:inc-gan}
	\end{figure}

	\subsubsection{Multi-scale Inception Feature Extraction}
	
\begin{figure}[!htb]\centering
		\vspace*{-0.5cm}
		
		\begin{tikzpicture}
			\node[] at(0,0){\includegraphics[scale=0.3,clip,trim={0cm 0cm 0cm 0cm}]{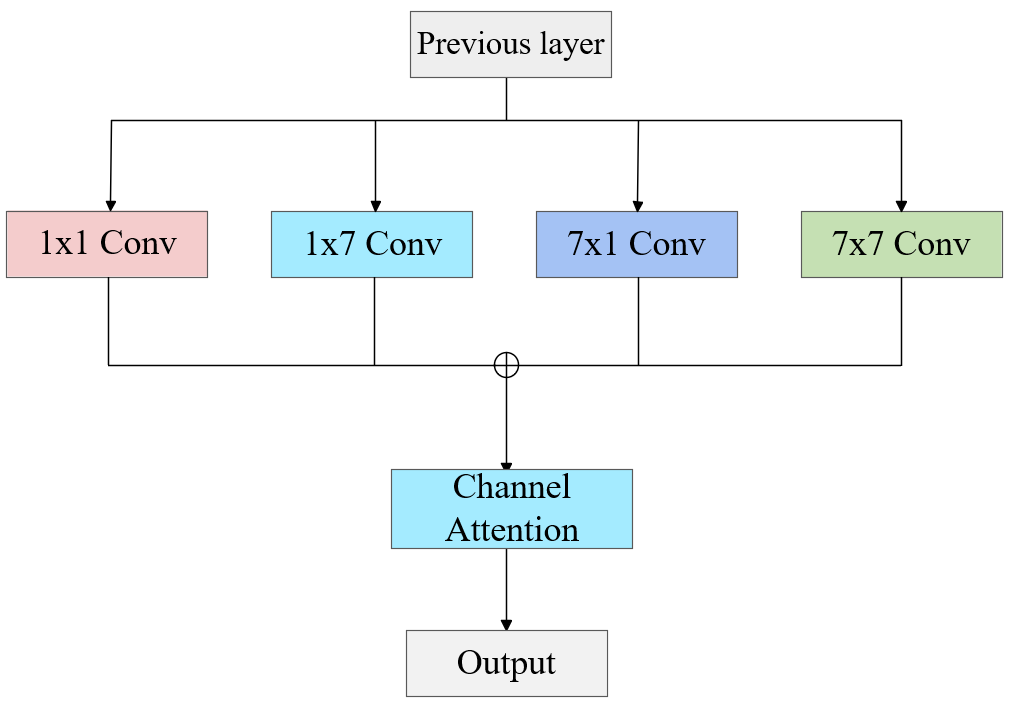}};
		\end{tikzpicture}
		\vspace*{-0.5cm}
		\caption{The structure of Inception block.}
		\label{fig: Inception Block}
	\end{figure}

	Radio wave propagation in tunnels exhibits multi-scale characteristics: macroscopic long-distance attenuation and microscopic local variations from wall reflections and scattering. To capture these features, we design Inception-style \cite{Laci25Deep, Szegedy16Inception} modules with four parallel branches (Fig.~\ref{fig: Inception Block}): $1 \times 1$ convolution for positional features, $1 \times 7$ and $7 \times 1$ convolution for tunnel width/height propagation, and $3 \times 3$ convolution for local spatial correlations.

	To adaptively learn the importance of different branch features, the module integrates a Channel Attention Mechanism (CAM)\cite{Wang20ECANet}:
	\vspace*{-0.1cm}
	\begin{equation}
		\text{CAM}(\mathbf{F}) = \sigma(\mathbf{W}_2 \cdot \text{ReLU}(\mathbf{W}_1 \cdot \text{GAP}(\mathbf{F})))
		\label{eq:attention}
	\end{equation}
	
	where $\mathbf{F}$ is the input feature, $\text{GAP}$ denotes global average pooling, $\mathbf{W}_1$ and $\mathbf{W}_2$ are learnable parameters, and $\sigma$ is the sigmoid function. This mechanism can dynamically adjust branch weights according to different tunnel geometries, significantly improving model adaptability.
	
	\subsubsection{Generator Network Design}
	
	The generator adopts a U-Net\cite{Ronneberger15UNet} architecture specifically optimized for processing 2-channel input (mask channel + one-channel data). The encoder path progressively extracts multi-scale features through multiple Inception blocks, while the decoder path combines skip connections to reconstruct complete electric field distributions.
	
	Unlike traditional square image inputs, this letter inputs rectangular tunnel cross-section images. Therefore, to support tunnel images of different resolutions, network depth employs adaptive adjustment:
	\begin{equation}
		N_{\text{layers}} = \min(5, \lfloor\log_2(\min(H, W))\rfloor - 2)
		\label{eq:adaptive_depth}
	\end{equation}
	
	where $H$ and $W$ represent image height and width respectively. The output layer uses ReLU activation functions to ensure physical non-negativity of electric field strengths.
\begin{table}[!t]
	\caption{Inc-GAN Parameters}
	\label{tab:network_parameters}
	\centering
	\begin{tabular}{|c|c|c|c|}
		\hline
		\textbf{Layer / Features} & \textbf{Channels} & \textbf{Layer / Features} & \textbf{Channels} \\
		\hline
		\multicolumn{2}{|c|}{\textbf{Encoder (Downsampling)}} & \multicolumn{2}{|c|}{\textbf{Decoder (Upsampling)}} \\
		\hline
		Input image & 1 & Output layer & 1 \\
		\hline
		Input processing & 64 & InceptionUp block$_3$ & 64 \\
		\hline
		Inception block$_1$ & 64 & Skip connection$_3$ & - \\
		\hline
		Downsample$_1$ & 64 & InceptionUp block$_2$ & 128 \\
		\hline
		Inception block$_2$ & 128 & Skip connection$_2$ & - \\
		\hline
		Downsample$_2$ & 128 & InceptionUp block$_1$ & 256 \\
		\hline
		Inception block$_3$ & 256 & Skip connection$_1$ & - \\
		\hline
		Downsample$_3$ & 256 & Generated image & - \\
		\hline
		\multicolumn{4}{|c|}{\textbf{Bottleneck}} \\
		\hline
		\multicolumn{2}{|c|}{Bottleneck block$_1$} & \multicolumn{2}{|c|}{256} \\
		\hline
		\multicolumn{2}{|c|}{Bottleneck block$_2$} & \multicolumn{2}{|c|}{256} \\
		\hline
		\multicolumn{4}{|c|}{\textbf{Discriminator Network}} \\
		\hline
		Disc layer$_1$ & 32 & Disc layer$_2$ & 64 \\
		\hline
		Disc layer$_3$ & 128 & Disc output & 1 \\
		\hline
	\end{tabular}
\end{table}

	\subsubsection{Discriminator Design}
	
\begin{figure}[htb]\centering
	\vspace*{-0.5cm}
	\begin{tikzpicture}
		\node[] at(0,0){\includegraphics[scale=0.5,clip,trim={0cm 0cm 0cm 0cm}]{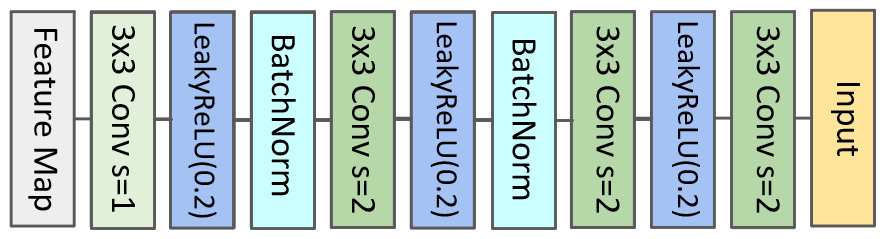}};
	\end{tikzpicture}
	\vspace*{-0.5cm}
	
	\caption{The architecture of discriminator.}\centering
	\label{fig: Discriminator}
\end{figure}

	As shown in Fig.~\ref{fig: Discriminator}, the discriminator adopts a PatchGAN architecture, improving generation quality through multi-scale local discrimination. All convolutional layers apply spectral normalization to enhance training stability, which is crucial for numerical stability in electric field reconstruction.
	
	\vspace*{-0.4cm}
	\subsection{Physics-Constrained Loss Functions}
	
	To ensure generated electric field distributions comply with physical laws, this letter designs specialized physics-constrained loss functions:
	
	Non-negativity constraint: Ensures electric field strength must be non-negative.
	\begin{equation}
		\mathcal{L}_{\text{non-neg}} = \mathbb{E}[\text{ReLU}(-\hat{\mathbf{E}})]
		\label{eq:loss_nonneg}
	\end{equation}
	
	Boundary condition constraint: Signal attenuation at tunnel boundaries.
	\begin{equation}
		\mathcal{L}_{\text{boundary}} = \mathbb{E}[\hat{\mathbf{E}}_{\text{boundary}}]
		\label{eq:loss_boundary}
	\end{equation}
	
	Spatial continuity constraint: Prevents non-physical discontinuities.
	\begin{equation}
		\mathcal{L}_{\text{smooth}} = \mathbb{E}[|\nabla_x \hat{\mathbf{E}}|] + \mathbb{E}[|\nabla_y \hat{\mathbf{E}}|]
		\label{eq:loss_smooth}
	\end{equation}
	
	where $\hat{\mathbf{E}}$ represents the predicted electric field distribution, and $\nabla_x$ and $\nabla_y$ represent spatial gradient operators in $x$ and $y$ directions respectively.

	\vspace*{-0.2cm}
\subsection{Progressive Training Strategy}

We employ progressive training to adapt sparse operational inputs, gradually reducing line retention ratio $\rho(t)$ from initial to final values over training epochs according to Eq.~\eqref{eq:progressive}. This strategy improves model robustness under actual measurement constraints.
\vspace*{-0.2cm}
\begin{equation}
	\rho(t) = \rho_{\text{init}} - \frac{(\rho_{\text{init}} - \rho_{\text{final}}) \cdot t}{T_{\text{prog}}}
	\label{eq:progressive}
\end{equation}
\vspace{-0.8cm}
	
	\subsection{Overall Loss Function}
	
	The final generator loss function comprehensively considers adversarial loss, content fidelity, and physics constraints:
	\begin{equation}
		\begin{aligned}
			\mathcal{L}_G = &\lambda_{\text{adv}} \mathcal{L}_{\text{adv}} + \lambda_{L1} \mathcal{L}_{L1} + \lambda_{\text{MSE}} \mathcal{L}_{\text{MSE}} \\
			&+ \lambda_{\text{SSIM}} \mathcal{L}_{\text{SSIM}} + \lambda_{\text{physics}} \mathcal{L}_{\text{physics}} 
		\end{aligned}
		\label{eq:total_loss}
	\end{equation}

		\section{NUMERICAL RESULTS}

		This section compares the traditional CGAN with the proposed Inc-GAN model for tunnel electric field reconstruction tasks. The training process considers typical rectangular cross-section tunnel geometries with standard railway tunnel dimensions.

		\vspace*{-0.4cm}
\label{NUMERICAL RESULTS}
\begin{figure}[htb]
	\centering
	\begin{minipage}{0.5\textwidth}
		\subfloat[PWE.]{
			\includegraphics[width=0.3\textwidth]{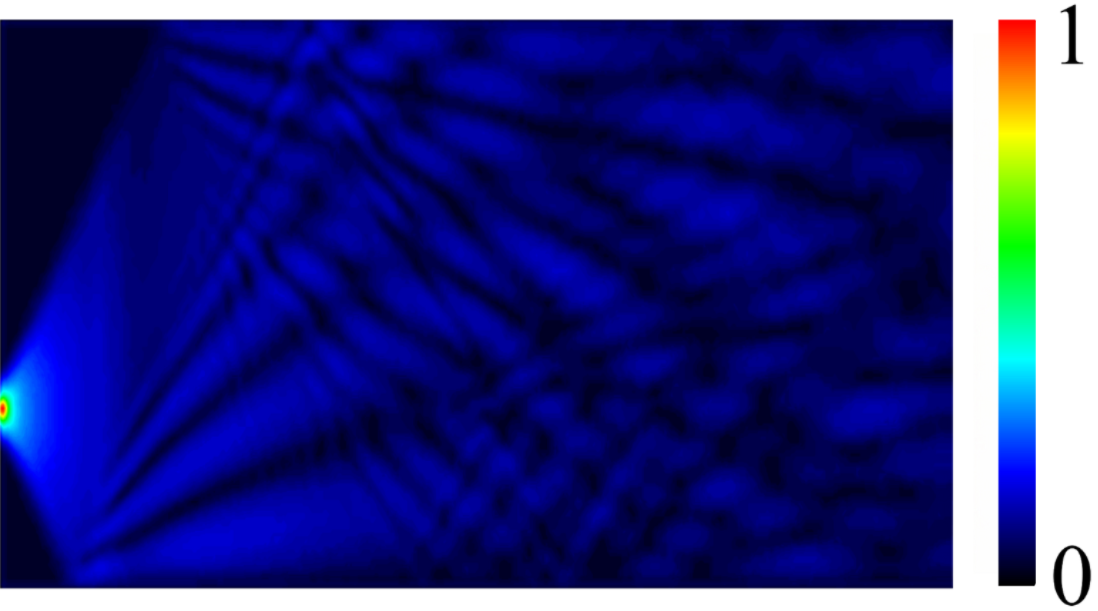}
		}
		\hfill
		\subfloat[CGAN Prediction.]{
			\includegraphics[width=0.3\textwidth]{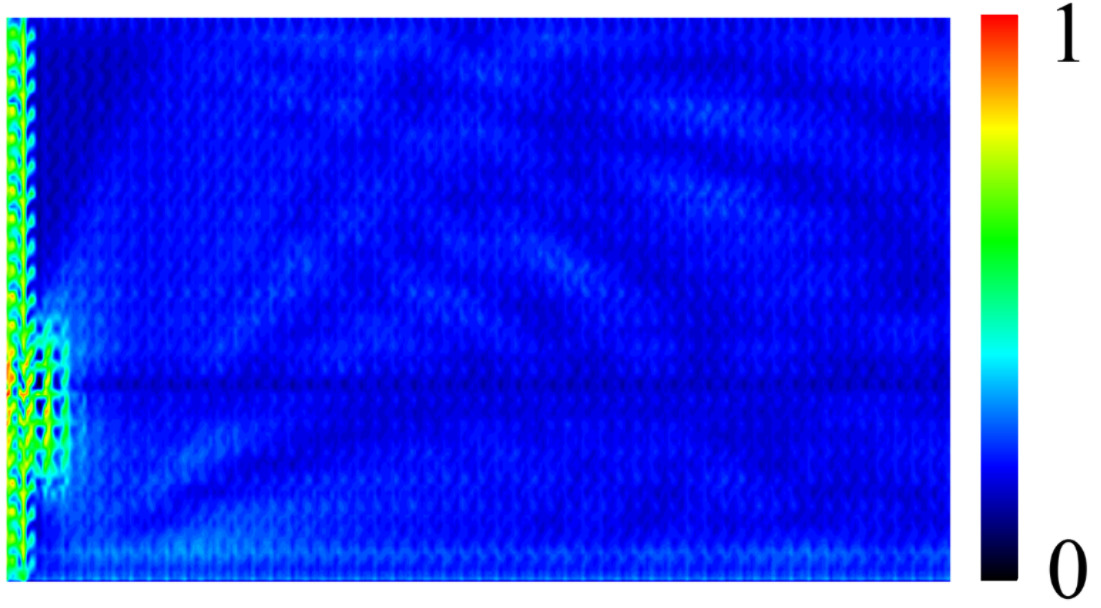}
		}
		\hfill
		\subfloat[Inc-GAN Prediction.]{
			\includegraphics[width=0.3\textwidth]{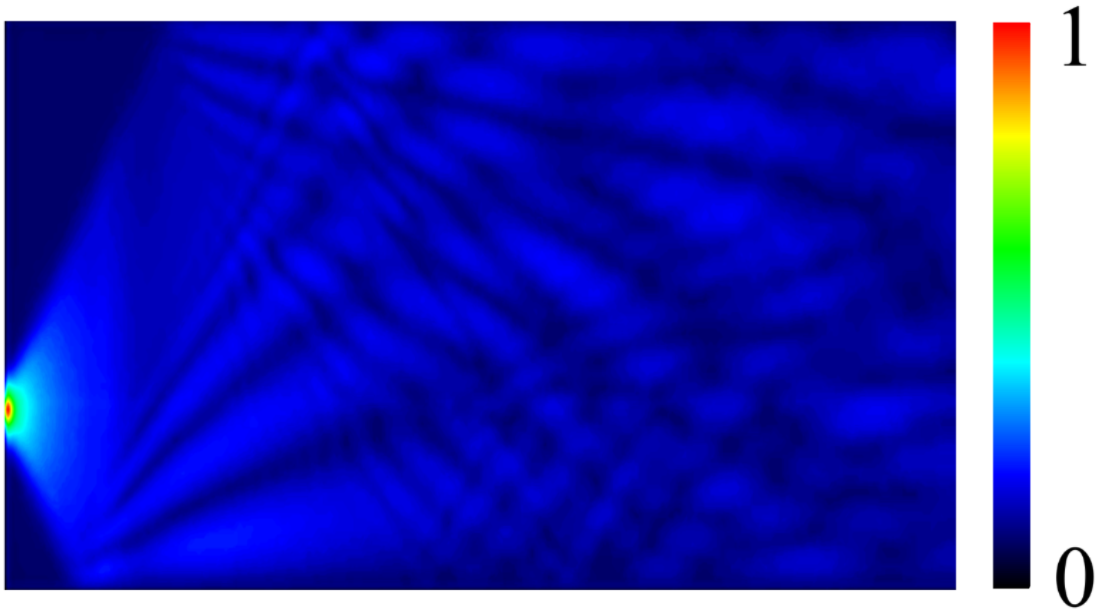}
		}\\[0.3em]
		
	\end{minipage}

	\caption{Comparison of electromagnetic field reconstruction results.}
	\label{fig:reconstruction_comparison}
\end{figure}

	\begin{figure*}[!htb]
	\centering
	\vspace*{-1.2cm}
	\subfloat[Inc-GAN RMSE: 0.007238 \\ CGAN RMSE: 0.15518]
	{
		\begin{tikzpicture}
			\node[] at(0,0){\includegraphics[scale=0.32,clip,trim={0cm 0cm 0cm 0}]{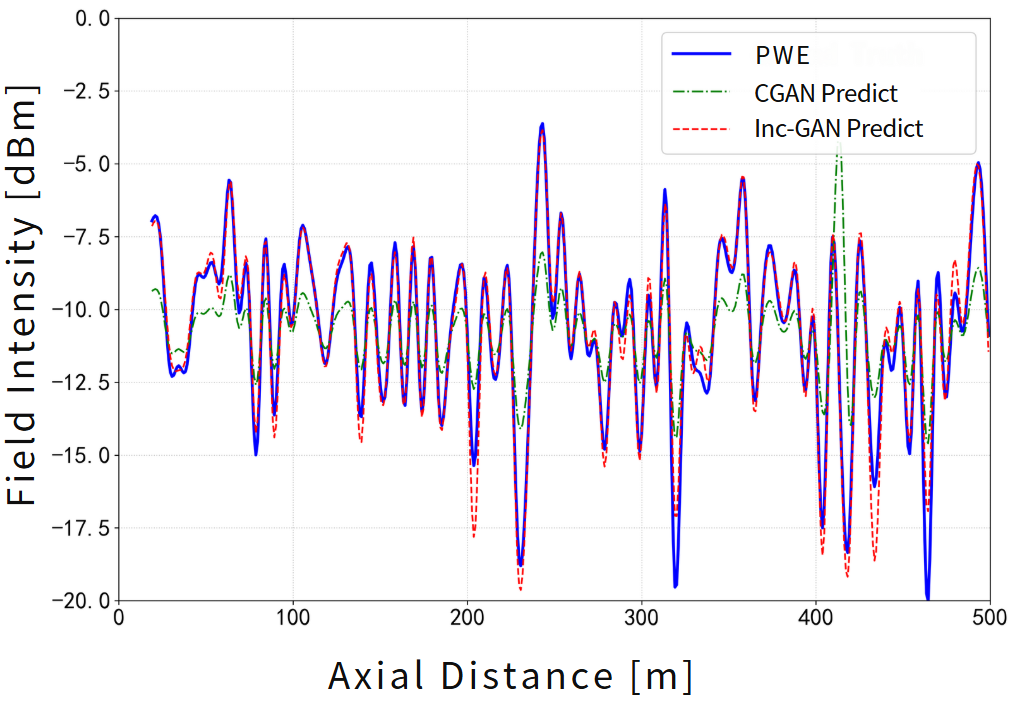}};
		\end{tikzpicture}
		\label{fig:rmse_a}
	}
	\hspace*{-0.6cm}
	\subfloat[ Inc-GAN RMSE: 0.006044 \\ CGAN RMSE: 0.12077]{
		\begin{tikzpicture}
			\node[] at(0,0){\includegraphics[scale=0.32,clip,trim={0cm 0cm 0cm 0cm}]{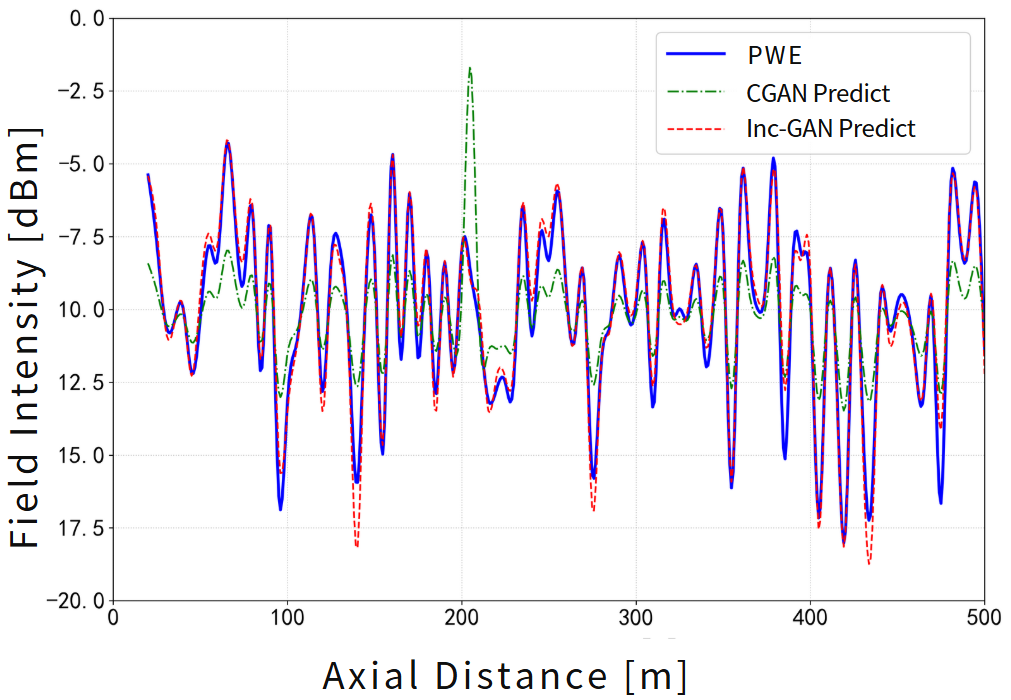}};
		\end{tikzpicture}
		\label{fig:rmse_b}
	}
\hspace*{-0.6cm}
	\subfloat[Inc-GAN MAE: 0.004247 \\ CGAN MAE: 0.09818]{
		\begin{tikzpicture}
			\node[] at(0,0){\includegraphics[scale=0.32,clip,trim={0cm 0.35cm 0cm 0.2cm}]{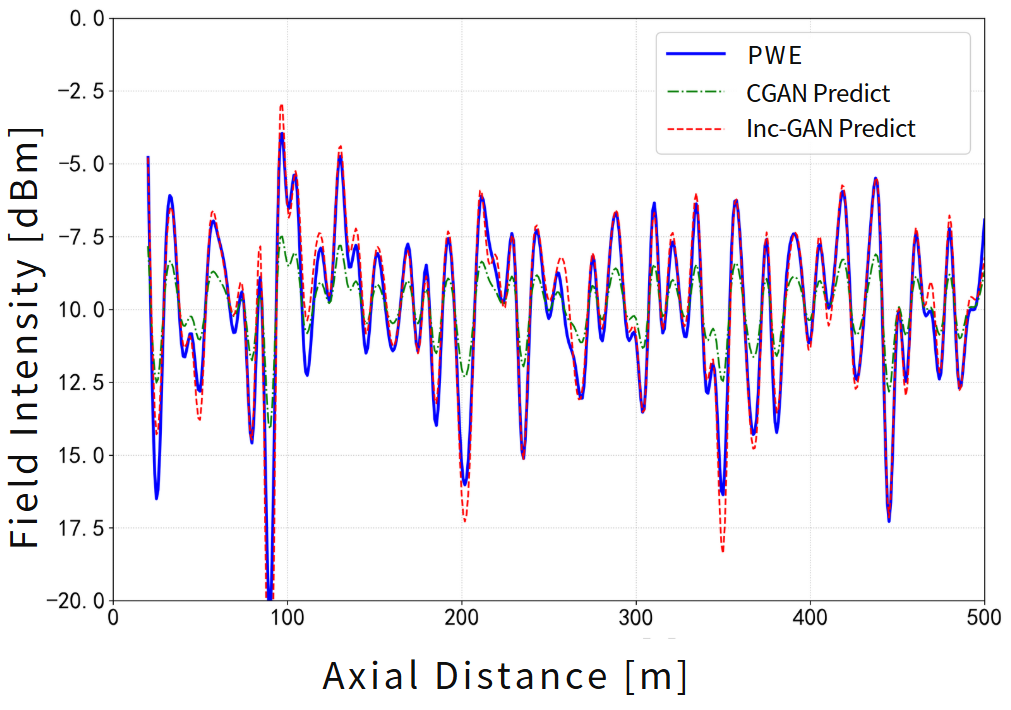}};
		\end{tikzpicture}
		\label{fig:mae_d}
	}

	\caption{Comparison between Inc-GAN and CGAN: Signal values from three randomly selected lines are compared against the Ground Truth.}
	\label{fig:training_comparison}
\end{figure*}

The environmental configuration uses rectangular tunnel cross-sections with \( 500 \,\text{m} \) length and \( 50 \,\text{m} \) height at spatial resolution of $1001 \times 101$, operating frequency range of \(0.9\,\text{--}\,5.8\,\text{GHz}\), relative permittivity $\varepsilon_r = 5$, relative permeability $\mu_r = 1$, and conductivity $\sigma \in [0.001, 0.1] \text{ S/m}
$. The input data consists of sparse line information extracted from complete electric field distributions, simulating the limited observational data available under actual train measurement conditions. The output target is complete two-dimensional electric field distribution reconstruction.
The spatial discretization step sizes used in PWE simulations are selected as $\Delta x = 0.5 \,\text{m}$ and $\Delta y = 1.0 \,\text{m}$, corresponding to a computational domain of 500 meters in length and approximately 100 meters in width.

Validation loss is measured through mean absolute error (MAE), root mean square error (RMSE)\cite{Wang23GRU,Sun20Exhaust}, and relative error, with specific formulas as follows:

\begin{align}
	\text{MAE} &= \frac{1}{N}\sum_{i=1}^{N}|y_i - \hat{y}_i| \\
	\text{RMSE} &= \sqrt{\frac{1}{N}\sum_{i=1}^{N}(y_i - \hat{y}_i)^2} \\
	\text{Rel. Error} &= \frac{\sqrt{\sum_{i=1}^{N}(y_i - \hat{y}_i)^2}}{\sqrt{\sum_{i=1}^{N}y_i^2}} \times 100\%
\end{align}

where $\hat{y}_i$ and $y_i$ represent electric field intensities obtained from Inc-GAN and PWE respectively, and $N$ is the total number of sampling points.

To validate the proposed network architecture, we selected traditional CGAN as the baseline model for conditional generation tasks. Both models used identical sparse input images, target electric field distributions, and implementation settings. Fig.~\ref{fig:training_comparison} compares Inc-GAN and CGAN prediction results for electric field reconstruction in test tunnels under different conditional line configurations. The proposed Inc-GAN model shows good consistency with actual PWE results, while CGAN exhibits obvious instability and larger errors.

Subsequently, we discuss the selection of physics-constrained loss weights. Higher physics constraint weights imply stronger physical consistency but may reduce reconstruction accuracy. Lower weights may lead to non-physical reconstruction results. To investigate this trade-off relationship, we trained four different models using four different physics constraint weights ($\gamma = 1, 5, 10, 20$). When physics constraint weights were 1, 5, 10, and 20, the models' average relative errors were 3.127\%, 2.804\%, 2.471\%, and 2.836\% respectively. Results show that when the physics constraint weight is 10, the model achieves optimal balance between accuracy and physical consistency.

In terms of computational efficiency, using the trained Inc-GAN model, inference time can be reduced to 0.04 seconds. In contrast, when using traditional PWE methods, the runtime for a single electric field calculation is approximately 91 seconds. 

\section{APPLICATIONS}
\label{APPLICATION:}
To validate the effectiveness of the proposed Inc-GAN model in practical engineering applications, this section applies it to electric field prediction in the Central Massif Tunnel in France\cite{Dudley07Tunnels}. This tunnel consists of large smooth stone blocks and concrete, with dimensions of \( 2500 \,\text{m} \) length, relative permittivity of 5, conductivity of \( 0.01 \,\text{S/m} \), and operating frequency of \( 900 \,\text{MHz} \)\cite{Dudley07Tunnels}, maintaining consistency with actual engineering environments. Considering the limitations of data acquisition during actual train operations, the model input uses signal source lines along the tunnel axis, which corresponds to sensor deployment constraints in engineering practice.

\begin{figure}[!t]
	\centering
	\includegraphics[width=0.4\textwidth]{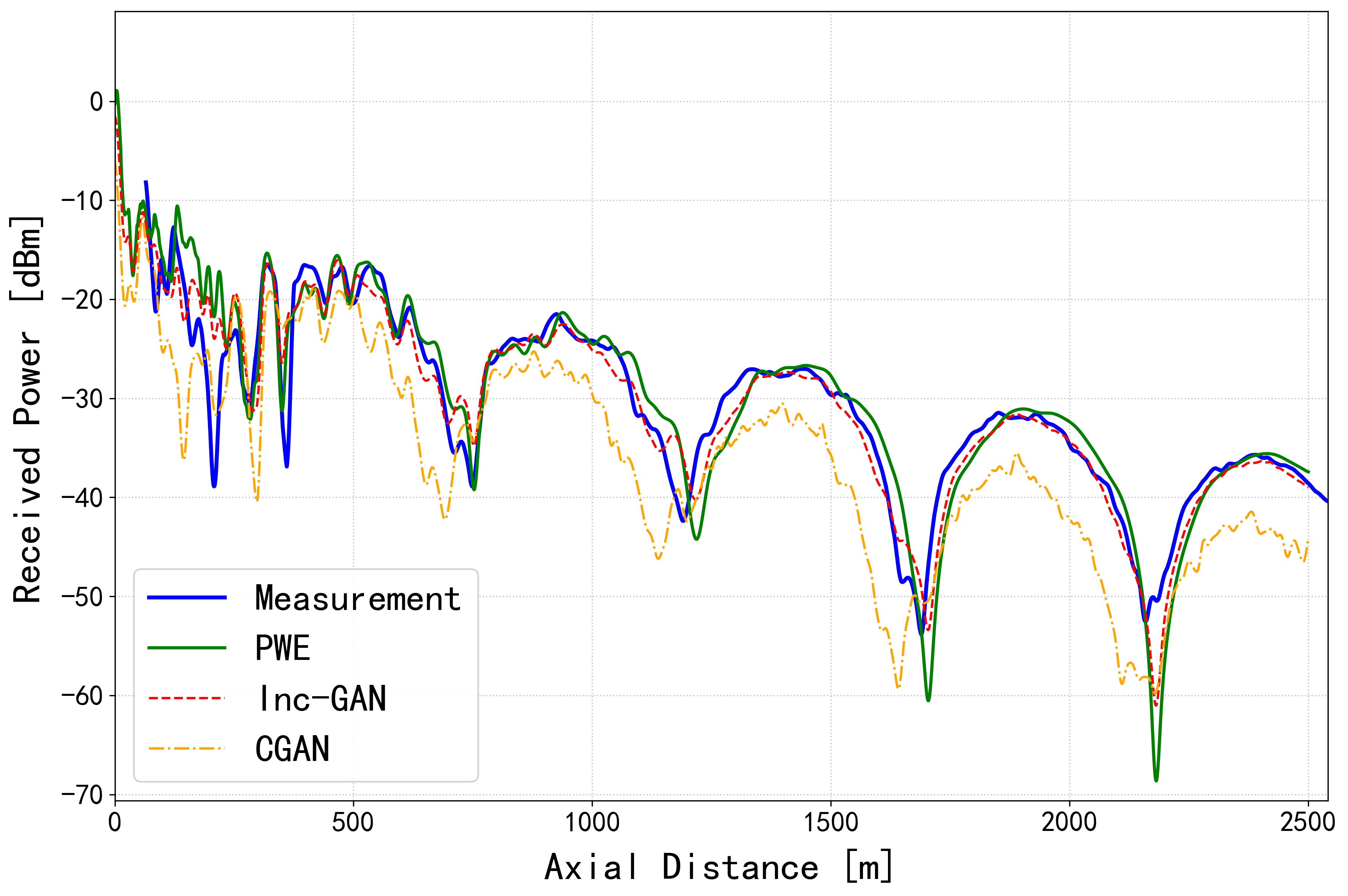}
	\vspace{-0.2cm}
	\caption{Comparison of received power predictions along the Central Massif Tunnel}
	\begin{tabular}{c}
		\footnotesize
	\end{tabular}
	\label{fig:tunnel_results}
\end{figure}

In Fig.~\ref{fig:tunnel_results}, we compare the electric field distribution prediction results generated by various models. Similar to previous cases, it can be observed that compared to CGAN, the proposed Inc-GAN model's prediction data shows higher agreement with VPE and measured values. The RMSE between measured data and predicted values are: CGAN \( 8.2 \,\text{dBm} \), Inc-GAN \( 2.1 \,\text{dBm} \). Regarding predicted values compared to measured data, the MAE for CGAN and Inc-GAN are 4.5 and \( 1.47 \,\text{dBm} \) respectively, representing an accuracy improvement of 67.3\%.

\section{CONCLUSION}
\label{CONCLUSION}

This letter proposed an Inc-GAN electric field reconstruction model for tunnel environments, providing a viable solution to replace traditional electric field prediction methods based on incomplete data input. By integrating multi-scale Inception modules and utilizing physics-constrained loss functions, this model effectively learns the mapping relationship between sparse line measurements and complete electric field distributions, thereby achieving precise reconstruction. Our proposed framework maintains accuracy comparable to traditional PWE methods while significantly reducing requirements for input data completeness. Additionally, to validate model effectiveness, we conducted simulation experiments in real tunnel scenarios, comparing with traditional CGAN methods and measured data for verification. This efficient tunnel environment electric field reconstruction technique is expected to be widely applied in intelligent rail transit communication fields. Future research will further explore electric field reconstruction problems in more complex three-dimensional tunnel structures and dynamic environments.

\ifCLASSOPTIONcaptionsoff
\newpage
\fi

\bibliographystyle{IEEEtran}

\begin{thebibliography}{10}
	\providecommand{\url}[1]{#1}
	\csname url@samestyle\endcsname
	\providecommand{\newblock}{\relax}
	\providecommand{\bibinfo}[2]{#2}
	\providecommand{\BIBentrySTDinterwordspacing}{\spaceskip=0pt\relax}
	\providecommand{\BIBentryALTinterwordstretchfactor}{4}
	\providecommand{\BIBentryALTinterwordspacing}{\spaceskip=\fontdimen2\font plus
		\BIBentryALTinterwordstretchfactor\fontdimen3\font minus
		\fontdimen4\font\relax}
	\providecommand{\BIBforeignlanguage}[2]{{%
			\expandafter\ifx\csname l@#1\endcsname\relax
			\typeout{** WARNING: IEEEtran.bst: No hyphenation pattern has been}%
			\typeout{** loaded for the language `#1'. Using the pattern for}%
			\typeout{** the default language instead.}%
			\else
			\language=\csname l@#1\endcsname
			\fi
			#2}}
	\providecommand{\BIBdecl}{\relax}
	\BIBdecl
	
	
		\bibitem{Qin25UAV}
	H.~Qin, Z.~Wu, Y.~Liu, X.~Zhang, and X.~Zhang, 
	``Physics-based trajectory design for cellular-connected {UAV} in rainy environments based on deep reinforcement learning,'' 
	\emph{IEEE Trans. Intell. Transp. Syst.}, vol.~26, no.~7, pp. 10320--10335, 2025.
	
\bibitem{CHEN202578}
W.~Chen, B.~Ai, Y.~Sun, C.~Yu, B.~Zhang, and C.~Yuen, ``Advanced 6G wireless communication technologies for intelligent high-speed railways,'' \emph{High-speed Railway}, vol.~3, no.~1, pp.~78--92, 2025.



\bibitem{qin2023physics}
H.~Qin and X.~Zhang, 
``Physics-based wave propagation model assisted vehicle localisation in tunnels,'' 
\emph{IET Microw. Antennas Propag.}, vol.~17, no.~10, pp. 786--796, 2023.



	
	\bibitem{Zhou17Alternative}
	L.~Zhou, C.~Liao, X.~Xiong, Q.~Zhang, and D.~Zhang, ``An alternative direction decomposition scheme and error analysis for parabolic equation model,'' \emph{IEEE Trans. Antennas Propag.}, vol.~65, no.~5, pp.~2547--2557, May 2017.
	
	\bibitem{He18Wave}
	Z.~He, T.~Su, H.-C.~Yin, and R.-S.~Chen, ``Wave propagation modeling of tunnels in complex meteorological environments with parabolic equation,'' \emph{IEEE Trans. Antennas Propag.}, vol.~66, no.~12, pp.~6629--6634, Dec. 2018.
	\bibitem{Yee66}
	K.~S.~Yee, ``Numerical solution of initial boundary value problems involving Maxwell's equations in isotropic media,'' \emph{IEEE Trans. Antennas Propag.}, vol.~14, no.~3, pp.~302--307, May 1966. 
	
	\bibitem{Taflove05}
	A.~Taflove and S.~C.~Hagness, \emph{Computational Electrodynamics: The Finite-Difference Time-Domain Method}, 3rd ed. Boston, MA: Artech House, 2005.
	\bibitem{Qian2025}
	C.~Qian, I.~Kaminer, and H.~Chen, ``A guidance to intelligent metamaterials and metamaterials intelligence,'' \emph{Nat. Commun.}, vol.~16, p.~1154, 2025.
	
	\bibitem{Qin23_AWPL_SSSPE}
	H.~Qin and X.~Zhang, 
	``Efficient radio wave propagation modeling in tunnels with a sparse Fourier transform-based split-step parabolic equation method,'' 
	\emph{IEEE Antennas Wireless Propag. Lett.}, vol.~22, no.~10, pp. 2442--2446, 2023.
	
	\bibitem{Qin23comparative}
	H.~Qin and X.~Zhang, 
	``Comparative analysis of finite-difference and split-step based parabolic equation methods for tunnel propagation modelling,'' 
	\emph{IET Microw. Antennas Propag.}, vol.~18, no.~2, pp. 59--72, 2024.
	
	
	\bibitem{Levy00}
	M.~F.~Levy, \emph{Parabolic Equation Methods for Electromagnetic Wave Propagation}, London, UK: IET, 2000. 
	
	\bibitem{Qin25ToA}
	H.~Qin, X.~Zhang, W.~Hou, and X.~Zhang, 
	``Fast parabolic wave equation-based time-of-arrival estimation exploiting sparse matrices,'' 
	\emph{IEEE Antennas Wireless Propag. Lett.}, vol.~24, no.~3, pp. 661--665, 2025.
	
	\bibitem{qin2023high}
	H.~Qin, S.~Huang, and X.~Zhang, 
	``A high-accuracy deep back-projection {CNN}-based propagation model for tunnels,'' 
	\emph{IEEE Antennas Wireless Propag. Lett.}, vol.~23, no.~3, pp. 1015--1019, 2023.
	
	
	
\bibitem{Seretis22}
A.~Seretis and C.~D.~Sarris, 
``An overview of machine learning techniques for radiowave propagation modeling,'' 
\emph{IEEE Trans. Antennas Propag.}, vol.~70, no.~6, pp. 3970--3985, 2022.

	

	
	\bibitem{Seretis2020}
	A.~Seretis, X.~Zhang, K.~Zeng, and C.~D.~Sarris, ``Artificial neural network models for radiowave propagation in tunnels,'' \emph{IET Microw., Antennas Propag.}, vol.~14, no.~11, pp.~1198--1208, 2020.
	
		\bibitem{Huang2025}
	S.~Huang, H.~Qin, W.~Hou, X.~Zhang, and X.~Zhang, ``Generalizable physics-guided convolutional neural network for irregular terrain propagation,'' \emph{IEEE Trans. Antennas Propag.}, 2025.
	
	
	\bibitem{Mallik2022}
	M.~Mallik, A.~A.~Tesfay, B.~Allaert, R.~Kassi, E.~Egea-Lopez, J.-M.~Molina-Garcia-Pardo, J.~Wiart, D.~P.~Gaillot, and L.~Clavier, ``Towards outdoor electromagnetic field exposure mapping generation using conditional {GANs},'' \emph{Sensors}, vol.~22, no.~24, p.~9643, 2022.
	\bibitem{Ye24GAN}
	M.~Ye, X.~Liang, C.~Pan, Y.~Xu, M.~Jiang, and C.~Li, ``GAN based near-field channel estimation for extremely large-scale MIMO systems,'' \emph{arXiv preprint arXiv:2402.17281}, Feb. 2024.

	
		 \bibitem{Laci25Deep}
	H.~Laçi, K.~Sevrani, and S.~Iqbal, ``Deep learning approaches for classification tasks in medical X-ray, MRI, and ultrasound images: a scoping review,'' \emph{BMC Med. Imaging}, vol.~25, art.~156, May 2025.
	\bibitem{Bordbar2024}
	A.~Bordbar, L.~Aabel, C.~Häger, C.~Fager, and G.~Durisi, ``Deep-learning-based channel estimation for distributed MIMO with 1-bit radio-over-fiber fronthaul,'' arXiv preprint arXiv:2406.11325, 2024.
	

	\bibitem{Mirza14CGAN}
	M.~Mirza and S.~Osindero, ``Conditional generative adversarial nets,'' \emph{arXiv preprint arXiv:1411.1784}, Nov. 2014.
	 \bibitem{Szegedy16Inception}
	 C.~Szegedy, V.~Vanhoucke, S.~Ioffe, J.~Shlens, and Z.~Wojna, ``Rethinking the inception architecture for computer vision,'' in \emph{Proc. IEEE Conf. Computer Vision and Pattern Recognition (CVPR)}, Las Vegas, NV, USA, Jun. 2016, pp.~2818--2826.

	 
	 \bibitem{Wang20ECANet}
	 Q.~Wang, B.~Wu, P.~Zhu, P.~Li, W.~Zuo, and Q.~Hu, ``ECA-Net: Efficient channel attention for deep convolutional neural networks,'' in \emph{Proc. IEEE/CVF Conf. Computer Vision and Pattern Recognition (CVPR)}, Seattle, WA, USA, Jun. 2020, pp.~11531--11539.
	 
	 \bibitem{Ronneberger15UNet}
	 O.~Ronneberger, P.~Fischer, and T.~Brox, ``U-Net: Convolutional networks for biomedical image segmentation,'' in \emph{Proc. Int. Conf. Medical Image Computing and Computer-Assisted Intervention (MICCAI)}, Munich, Germany, Oct. 2015, pp.~234--241.
	 
	 
	 
	 \bibitem{Wang23GRU}
	 S.~Wang, K.~Yang, Y.~Shi, F.~Yang, H.~Zhang, and Y.~Ma, ``Prediction of over-the-horizon electromagnetic wave propagation in evaporation ducts based on the gated recurrent unit network model,'' \emph{IEEE Trans. Antennas Propag.}, vol.~71, no.~4, pp.~3485--3496, Apr. 2023.
	 
	 \bibitem{Sun20Exhaust}
	 B.~Sun, K.~Xie, L.~Shi, M.~Yang, B.~Yao, and S.~Guo, ``Experimental investigation on electromagnetic waves transmitting through exhaust plume: From propagation to channel characteristics,'' \emph{IEEE Trans. Antennas Propag.}, vol.~68, no.~12, pp.~8021--8032, Dec. 2020.
	 
	 \bibitem{Dudley07Tunnels}
	 D.~G.~Dudley, M.~Lienard, S.~F.~Mahmoud, and P.~Degauque, ``Wireless propagation in tunnels,'' \emph{IEEE Antennas Propag. Mag.}, vol.~49, no.~2, pp.~11--26, Apr. 2007.
\end{thebibliography}

\end{document}